\documentclass[aps,prl,twocolumn,showpacs,superscriptaddress]{revtex4-1}
\usepackage{graphicx}
\usepackage{amsmath}
\usepackage{amssymb}
\usepackage{colordvi}
\usepackage{mathrsfs}
\usepackage{bm}
\usepackage{verbatim}
\usepackage{dcolumn}
\usepackage{bm}
\usepackage{epsfig}
\usepackage{subfigure}

\begin{document}
\title{Pursuing High-Temperature Quantum Anomalous Hall Effect in MnBi$_2$Te$_4$/Sb$_2$Te$_3$ Heterostructures}
\author{Shifei Qi}
\affiliation{ICQD, Hefei National Laboratory for Physical Sciences at Microscale, CAS Key Laboratory of Strongly-Coupled Quantum Matter Physics, and Department of Physics, University of Science and Technology of China, Hefei, Anhui 230026, China}
\affiliation{Department of Physics, Hebei Normal University, Shijiazhuang, Hebei 050024, China}
\author{Ruiling Gao}
\affiliation{Institute of Materials Science, Shanxi Normal University, Linfen, Shanxi 041004, China}
\affiliation{Department of Physics, Hebei Normal University, Shijiazhuang, Hebei 050024, China}
\author{Maozhi Chang}
\affiliation{Institute of Materials Science, Shanxi Normal University, Linfen, Shanxi 041004, China}
\affiliation{Department of Physics, Hebei Normal University, Shijiazhuang, Hebei 050024, China}
\author{Yulei Han}
\affiliation{ICQD, Hefei National Laboratory for Physical Sciences at Microscale, CAS Key Laboratory of Strongly-Coupled Quantum Matter Physics, and Department of Physics, University of Science and Technology of China, Hefei, Anhui 230026, China}
\author{Zhenhua Qiao}
\email[Correspondence author:~~]{qiao@ustc.edu.cn}
\affiliation{ICQD, Hefei National Laboratory for Physical Sciences at Microscale, CAS Key Laboratory of Strongly-Coupled Quantum Matter Physics, and Department of Physics, University of Science and Technology of China, Hefei, Anhui 230026, China}
\date{\today{}}
\begin{abstract}
  Quantum anomalous Hall effect (QAHE) has been experimentally realized in magnetically-doped topological insulators or intrinsic magnetic topological insulator MnBi$_2$Te$_4$ by applying an external magnetic field. However, either the low observation temperature or the unexpected external magnetic field (tuning all MnBi$_2$Te$_4$ layers to be ferromagnetic) still hinders further application of QAHE. Here, we theoretically demonstrate that proper stacking of MnBi$_2$Te$_4$ and Sb$_2$Te$_3$ layers is able to produce intrinsically ferromagnetic van der Waals heterostructures to realize the high-temperature QAHE. We find that interlayer ferromagnetic transition can happen at $T_{\rm C}=42~\rm K$ when a five-quintuple-layer Sb$_2$Te$_3$ topological insulator is inserted into two septuple-layer MnBi$_2$Te$_4$ with interlayer antiferromagnetic coupling. Band structure and topological property calculations show that MnBi$_2$Te$_4$/Sb$_2$Te$_3$/MnBi$_2$Te$_4$ heterostructure exhibits a topologically nontrivial band gap around 26 meV, that hosts a QAHE with a Chern number of $\mathcal{C}=1$. In addition, our proposed materials system should be considered as an ideal platform to explore high-temperature QAHE due to the fact of natural charge-compensation, originating from the intrinsic n-type defects in MnBi$_2$Te$_4$ and p-type defects in Sb$_2$Te$_3$.
\end{abstract}

\maketitle
\textit{Introduction---.}
As a typical topological state, quantum anomalous Hall effect (QAHE) has been firstly observed in magnetic-element doped topological insulator (TI) thin films~\cite{ChangCuiZu,ExperimentalQAHE1,ExperimentalQAHE2,2015NatMat,APL}. Due to its robustness against local perturbations, the corresponding chirally propagating edge mode has immense potential in future low-power or dissipationless electronic device applications~\cite{weng2015quantum,ren2016topological,he2018topological}. QAHE was initially predicated in a honeycomb lattice toy model by F. M. Haldane in 1988~\cite{Haldane}. After then, much effort has been made to exploring new systems for realizing the QAHE in these related systems~\cite{proposal1,proposal2,proposal3,proposal4,proposal5,proposal6,proposal7,proposal8,proposal9,proposal10,proposal11,proposal12}. Because of the inherently strong spin-orbit coupling in $\mathbb{Z}_2$ topological insulators ~\cite{TopologicalInsulator1,TopologicalInsulator2}, it was predicted that ferromagnetic TI thin films can give rise to QAHE. One convenient approach to induce magnetism in TIs is to dope magnetic elements~\cite{DMS,proposal4,TI-magnetism1,TI-magnetism2,TI-magnetism3, Qi2016,CodopingEXP}. This has indeed led to the first experimental observation of QAHE~\cite{ChangCuiZu}. So far, all the experimentally observed QAHEs in doped TIs were achieved at extremely low temperatures, typically $\sim$30 mK, making drastically increasing the QAHE observation temperature a daunting challenge both fundamentally and for potential applications.

Aside from the magnetic-element doped TIs,  sandwiched TI heterostructures between two ferromagnetic insulator (FMI) layers were considered as another practical platform, because of the more ``clean" ferromagnetism with higher Curie temperature than that in the magnetic-element doped TIs~\cite{HS-Rev,Heterostructure1}. Unfortunately, due to the poor interfaces, the observed anomalous Hall resistance cannot reach the quantized value, though several FMI/TI heterostructures have been successfully fabricated in experiments~\cite{ExpHS1,ExpHS2,ExpHS3,ExpHS4,ExpHS5,ExpHS6,ExpHS7}. Considering the better control of interfaces, van der Waals (vdW) layered magnetic materials exhibit great advantage in designing high-quality FMI/TI heterostructures.

\begin{figure}
  \includegraphics[width=8.0cm,angle=0]{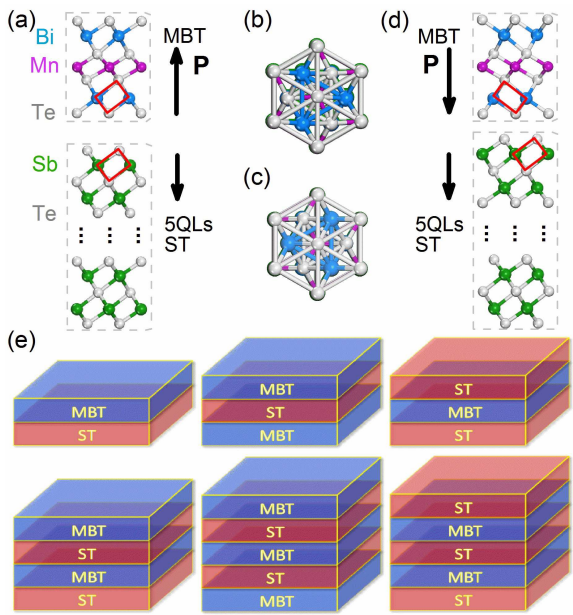}
  \caption{(a) Side view and (b) top view of crystal structures of ABC-ABC stacking heterostructure constructed by 1SL MnBi$_2$Te$_4$ and 5QL Sb$_2$Te$_3$; (c) The corresponding top view and (d) side view of crystal structure of ABC-ACB stacking case. Arrows (P) indicate the direction of spontaneous electric polarization. Red-solid frames in panels (a) and (d) label different stacking of 1SL MnBi$_2$Te$_4$ and 5QL Sb$_2$Te$_3$. (e) Schematics of six different heterostructures with MBT and ST representing MnBi$_2$Te$_4$ and Sb$_2$Te$_3$, respectively.}
\label{Fig1}
\end{figure}

Recently, MnBi$_2$Te$_4$ (vdW compounds) was successfully synthesized and confirmed as an intrinsic magnetic TI in experiments~\cite{MBT-HeK,MBT-XuY}. It possesses layered structure with a triangle lattice, as displayed in Fig.~\ref{Fig1}(a) and ~\ref{Fig1}(b). So far, rich topological states were proposed in MnBi$_2$Te$_4$ due to its characteristics of intralayer ferromagnetic and interlayer antiferromagnetic exchange interactions~\cite{ZhangHJ-PRL,MBT-HeK,MBT-XuY,ZhangHJ-2}. In particular, QAHE in MnBi$_2$Te$_4$ system has been observed at 4.5 K by applying an external magnetic field to convert the initial antiferromagnetism between layers to be ferromagnetism~\cite{MBT-QAHE}. This is an undoubtedly obvious enhancement of QAHE observation temperature. To realize QAHE without applying any external field in MnBi$_2$Te$_4$ systems, interlayer antiferromagnetic coupling is a big obstacle. It is known that single layer MnBi$_2$Te$_4$ is a trivial FMI. Thus, one can construct the vdW MnBi$_2$Te$_4$/TI heterostructures with atomically perfect interface and intrinsic magnetism. We noted that MnBi$_2$Te$_4$/Bi$_2$Te$_3$ (not a TI) heterostructures have been successfully fabricated in experiment~\cite{MBT-HS}. Although similar MnBi$_2$Te$_4$/Bi$_2$Te$_3$ heterostructure was theoretically predicted to be good candidate for realizing QAHE~\cite{MBT-HS-Theor}, the naturally existing n-type defects in both MnBi$_2$Te$_4$ and Bi$_2$Te$_3$ make their heterostructures metallic. On the contrary, considering the intrinsic p-type defects in Sb$_2$Te$_3$ (a 3D TI), MnBi$_2$Te$_4$/Sb$_2$Te$_3$ heterostructure should be charge compensated, which resolves the obstacle for realizing QAHE in MnBi$_2$Te$_4$ system.

In this Letter, we theoretically show that high-temperature QAHE can be realized in MnBi$_2$Te$_4$/Sb$_2$Te$_3$ heterostructure without additional magnetic field. We first demonstrate that an interlayer ferromagnetic transition can occur at $T_{\rm C}=42$~K when a five quintuple-layer (QL) Sb$_2$Te$_3$ is inserted between two septuple-layer (SL) MnBi$_2$Te$_4$ with interlayer antiferromagnetic coupling. Further calculation indicates that the sandwiched heterostructure is an intrinsic QAHE with a bulk gap of ~26 meV and a nonzero Chern number of $\mathcal{C}=1$. Our study not only reveals the possibility of producing ferromagnetic MnBi$_2$Te$_4$ to remove the unexpected external magnetic field in observing QAHE, but also provides an experiment-friendly sandwiched system to realize high-temperature QAHE.

\textit{Methods and Systems---.}
Our first-principles calculations were performed by using the projected augmented-wave method~\cite{PAW} as implemented in the Vienna Ab-initio Simulation Package (VASP)~\cite{VASP1}. The generalized gradient approximation (GGA) of Perdew-Burke-Ernzerhof (PBE) type was used to treat the exchange-correlation interaction~\cite{GGA}. A $1\times 1$ MnBi$_2$Te$_4$/Sb$_2$Te$_3$ supercell is chosen in our study. For thin film calculations, the Sb$_2$Te$_3$ film thickness was chosen to be 5QL. A vacuum buffer space of 20~\AA  was used to prevent the coupling between adjacent slabs. The kinetic energy cutoff was set to be 350 eV. During structural relaxation, all atoms were allowed to relax until the Hellmann-Feynman force on each atom is smaller than 0.01eV/\AA. The Brillouin-zone integration was carried out by using $9\times9\times1$ Monkhorst-Pack grids for different heterostructure systems. Unless mentioned otherwise, spin-orbit coupling and the GGA+U method were used with the on-site repulsion parameter U=3.00 and 5.34 eV~\cite{GGAU} and the exchange parameter J=0 eV, where U is for the more localized $3d$ orbitals of Mn atoms in all calculations. Different U values provide similar results in our calculations. In addition, the vdW interactions were considered in our numerical calculations~\cite{DFTD2}. The Curie temperature $T_{\rm C}$ was estimated within the mean-field approximation ${k_{\rm B}} {T_{\rm C}} = {2}/{3} J x$~\cite{CurieTemperature},where $k_{\rm B}$ is the Boltzmann constant, $x$ is the dopant concentration, and $J$ is the exchange parameter obtained from the total energy difference between ferromagnetic (FM) and antiferromagnetic (AFM) configurations in different heterostructures.

\begin{table*}
  \caption{\label{table1}Total magnetic moments, magnetic order, energy difference between FM and AFM configurations $\Delta E_{\rm FM-AFM}$, and Curie temperature $T_{\rm C}$ in 2SLs MnBi$_2$Te$_4$ and different MnBi$_2$Te$_4$ and Sb$_2$Te$_3$ heterostructures.}
  \begin{ruledtabular}
  \begin{tabular}{lcccc}
  Systems&Moments/$\mu_{B}$& Magnetic Order&$\Delta$E/meV&$T_{\rm C}$/K\\ \hline
  2SLs MBT           &$0.0$  & AFM  &$9.3$  &$-$ \\
  MBT/ST            &$5.0$   & FM   &$-$    &$14$\\
  MBT/ST/MBT        &$10.0$  & FM   &$-27.9$ &$42$\\
  ST/MBT/ST         &$5.0$   & FM   &$-$  &$14$\\
  MBT/ST/MBT/ST     &$0.0$  & AFM  &$0.3$  &$-$\\
  MBT/ST/MBT/ST/MBT &$5.0$   & AFM  &$2.4$ &$-$\\
  ST/MBT/ST/MBT/ST  &$0.0$  & AFM  &$11.4$&$-$ \\
 \end{tabular}
 \end{ruledtabular}
\end{table*}

MnBi$_2$Te$_4$ is composed of SL blocks stacked one by one along the [0001] direction via vdW interaction~\cite{MBT-HeK,MBT-XuY} [see Fig.~\ref{Fig1}(a)]. The resulting magnetic order appears to be interlayer antiferromagnetic, where the ferromagnetic Mn layers of neighboring blocks are coupled in an antiparallel manner~\cite{MBT-HeK,MBT-XuY}. Among various 3D TIs, Bi$_2$Te$_3$ should be the most optimal candidate to construct MnBi$_2$Te$_4$/TI heterostructures due to the small lattice mismatch (1.3\%). However, as observed in related experiments, MnBi$_2$Te$_4$ and Bi$_2$Te$_3$ are respectively electron conducting due to their intrinsic n-type defects. Thus, their heterostructures are metallic (i.e., the Fermi level is out of the band gap). Considering the intrinsic p-type conducting character of Sb$_2$Te$_3$ (lattice mismatch is about 2.3\%), the MnBi$_2$Te$_4$/Sb$_2$Te$_3$ heterostructure can be tuned to be charge compensated, which is an important prerequisite condition for realizing QAHE. Therefore, in our work, we focused on different heterostructures [See Fig.~\ref{Fig1}(e)] constructed by MnBi$_2$Te$_4$ and Sb$_2$Te$_3$.

\textit{Structural Stability---.}
Let us first study the structural stability of heterostructures. It is noted that the stacking of MnBi$_2$Te$_4$ and Sb$_2$Te$_3$ is different from the similar system MnBi$_2$Te$_4$/Bi$_2$Te$_3$~\cite{MBT-HS-Theor}, where the ABC-ABC stacking is most favorable. Two different stacking heterostructures of MnBi$_2$Te$_4$ and Sb$_2$Te$_3$ are provided. One is in the ABC-ABC stacking [Fig.~\ref{Fig1}(a)], and the other is in the ABC-ACB stacking [Fig.~\ref{Fig1}(d)]. Their top views are respectively displayed in Figs.~\ref{Fig1}(b) and \ref{Fig1}(c).  Our numerical calculations show that the ABC-ACB stacking is more preferred based on the calculated result that the total energy of ABC-ACB stacking structure is -30.8 meV/unit cell lower than that of ABC-ABC stacking structure. The underlying reason is closely related to the dipole-dipole interaction between MnBi$_2$Te$_4$ and Sb$_2$Te$_3$. As highlighted in Figs.~\ref{Fig1}(a)and \ref{Fig1}(d), the dipole-moment directions of MnBi$_2$Te$_4$ and Sb$_2$Te$_3$ are opposite in the ABC-ABC stacking case, while they are the same in the ABC-ACB stacking one. All these indicate that the dipole-dipole interactions are respectively repulsive and attractive in ABC-ABC and ABC-ACB stacking. This finding can also be reflected by the larger interfacial distance in the ABC-ABC stacking than that in the ABC-ACB stacking [see Figs.~\ref{Fig1}(a) and \ref{Fig1}(d)]. According to these finding, hereinbelow, we adopted the ABC-ACB stacking system to further investigate the magnetic properties, band structures, and topological properties.

\textit{Magnetic Properties---.}
As displayed in Tab.~\ref{table1}, we first confirm the magnetic properties of 2SLs MnBi$_2$Te$_4$. From our calculations, the 2SLs film indeed exhibits interlayer antiferromagnetism (i.e., 9.3 meV lower than that in ferromagnetic order). The local magnetic moment in 1SL is $5\mu_{B}$, agreeing with previous report~\cite{MBT-XuY}, even if it is inserted into different heterostructures. Among all six studied heterostructures, only three of them are in the ferromagnetic ordering state. Especially for the MBT/ST/MBT sandwiched system, we found that a ferromagnetic phase transition occurs when a 5QL Sb$_2$Te$_3$ is inserted between two 2SL MnBi$_2$Te$_4$ thin films with interlayer antiferromagnetic coupling. This finding agrees well with a recent experimental result~\cite{MBT-HS}, i.e., the interlayer antiferromagnetic exchange interaction can be gradually weakened with increasing the separation between ferromagnetic layers in the MnBi$_2$Te$_4$/Bi$_2$Te$_3$ heterostructure and a ferromagnetic state could be established below 5K~\cite{MBT-HS}. The total energy difference between FM and AFM states is -27.9 meV in the MBT/ST/MBT system and the estimated Curie temperature $T_{\rm C}$ is around 42K. The other favorable FM state can also be obtained in MBT/ST, ST/MBT/ST, and odd-SL MBT/ST/MBT/ST/MBT systems, which are highly desirable candidates for realizing high-temperature QAHE. In fact, the magnetic ground states of MBT/ST, MBT/ST/MBT, and ST/MBT/ST heterostructure have been calculated with the inclusion of spin-orbit coupling, and the results indicate that out-of-plane magnetic structures are more preferred, which are respectively about 0.01, 0.06, and 0.03 meV(per Mn atom) lower than those of in-plane magnetic structures of MBT/ST, MBT/ST/MBT, and ST/MBT/ST heterostructures.

\begin{figure*}
  \includegraphics[width=12cm,angle=0]{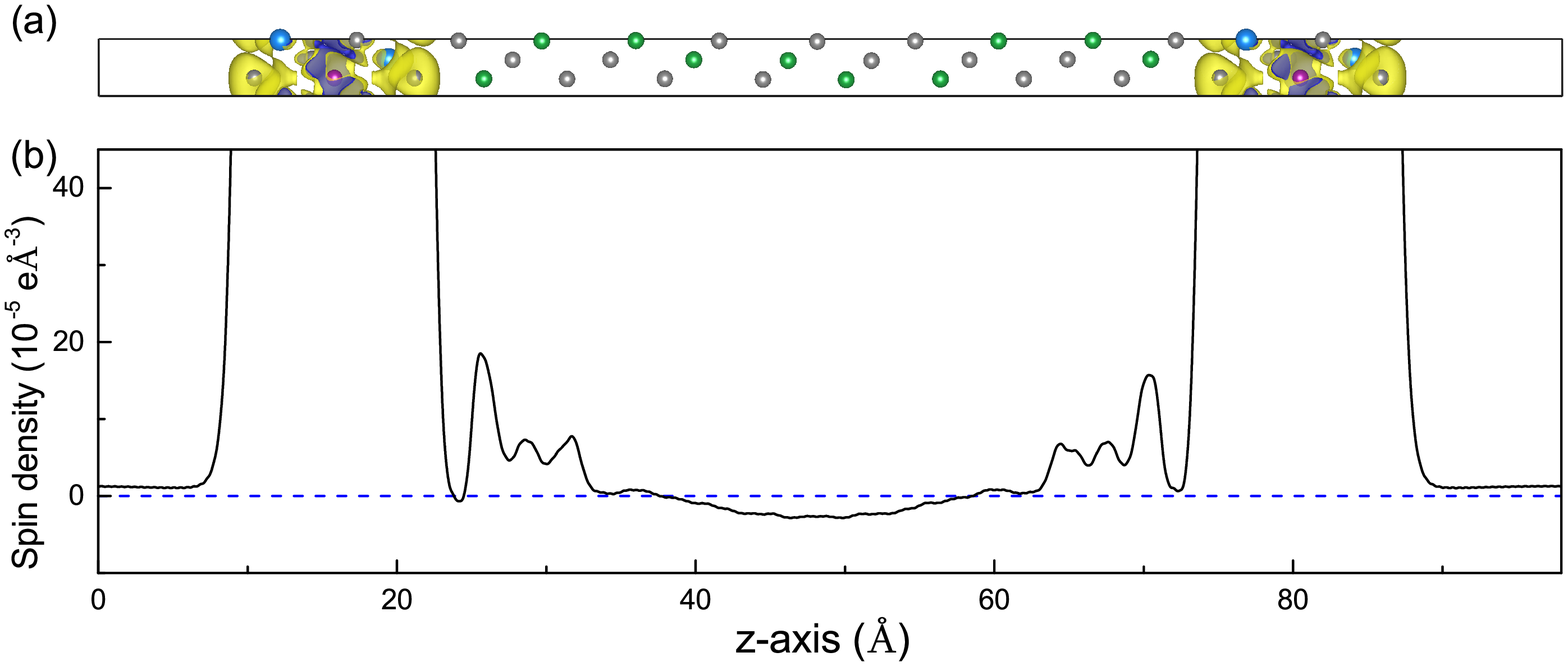}
  \caption{(a) The space distribution of spin density of MBT/ST/MBT heterostructure. The yellow and blue isosurfaces in the image respectively represent majority and minority spin.(b) The layer averaged spin density along the z direction of MBT/ST/MBT heterostructure.} \label{Fig2}
\end{figure*}

For the TI/magnetic insulators heterostructures, surface carrier-mediated interaction is found to be responsible for long-range ferromagnetism. In especially, Kou et al. found that the penetration depth of the surface states is a typical value of 2-3 QLs, and then they observe that fewer surface Dirac fermions will participate into the RKKY mediating process with larger TI layer depth, therefore weaking the surface-related RKKY coupling strength~\cite{Kou}. For our MBT/ST heterostructures, surface carrier-mediated interaction is responsible for the long-range ferromagnetism. In the MBT/ST/MBT system, as shown in the Figs.~\ref{Fig2}(a), the FM spin polarizations mainly appear at the interfaces of MBT and ST topological insulator and long-range FM coupling exists in this system. According to Kou et al's finding, the penetration depth of the surface states is 2-3 QLs. For 5QLs ST topological insulator, the penetration depths of surface states of the top and bottom layers are all 2-3 QLs. The Figs.~\ref{Fig2}(b) implies exchange interactions between different QLs. Hence, long-range FM coupling can be realized by surface carriers in the MBT/ST/MBT system.

For the MBT/ST/MBT/ST/MBT system, there should be three possible magnetic states including AFM1($\uparrow$$\uparrow$$\downarrow$), AFM2 ($\uparrow$$\downarrow$$\uparrow$), and FM ($\uparrow$$\uparrow$$\uparrow$) where only magnetic moments of MBT are shown. From our calculated results, AFM1 is the most stable one, then the relative energies of FM and AFM2 states are respectively 2.4 and 3.6 meV higher than that of AFM1 state. Hence, the magnetic property and band structure of MBT/ST/MBT/ST/MBT system are corresponding to the AFM1 state. In the AFM1, the MBT/ST/MBT part of MBT/ST/MBT/ST/MBT system is still FM coupling, only loner range FM coupling do not realized (FM state). This is because that longer range FM coupling (FM state) is not only dependent on surface carriers in one ST unit mediation, but also surface carriers from the first and the second ST units of the MBT/ST/MBT/ST/MBT system. The total penetration depth of surface states of the top layer of the first ST unit and the bottom layer of the second layer is only 4-6 QLs, which is smaller than distance between the first MBT and the third MBT. Hence FM state is not the most stable one for the MBT/ST/MBT/ST/MBT system.

\begin{figure}
  \includegraphics[width=8cm,angle=0]{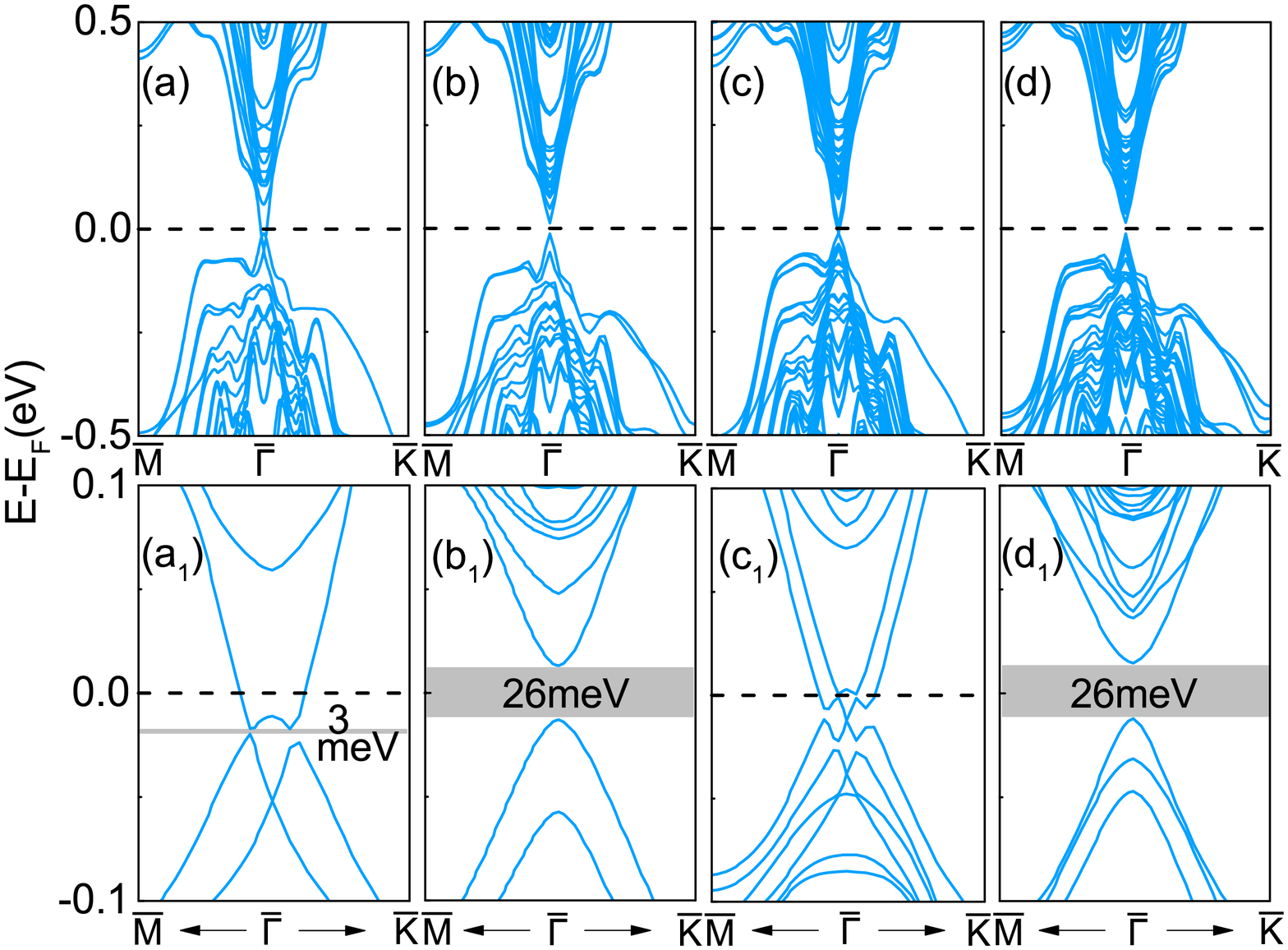}
  \caption{Band structures of four different ferromagnetic heterostructures along high-symmetry lines. (a)-(d) are respectively for MBT/ST, MBT/ST/MBT, ST/MBT/ST, and MBT/ST/MBT/ST/MBT. (a1)-(d1) are the corresponding Zooming-out near Fermi level. The dashed line denotes the Fermi level.} \label{Fig3}
\end{figure}

\textit{Band Structures and QAHE---.}
To confirm that such systems can indeed realize QAHE, band structures of these heterostructures are necessary. First, we presented that bulk band gaps can be opened up around the Dirac point $\Gamma$, which in principle guarantees the insulating state~\cite{TopologicalInsulator2}. Figure~\ref{Fig3} displays the band structures along high-symmetry lines for four different heterostructures. One can see that for MBT/ST system, the opened band gap is only 3 meV [see panel (a)]; while for MBT/ST/MBT [see panel (b)] and odd-SL MBT/ST/MBT/ST/MBT [see panel (d)] systems, the same sizeable band gaps of 26 meV are opened; and for ST/MBT/ST [see panel (c)], no band gap opens but a Dirac point is formed.

\begin{figure}
  \includegraphics[width=8cm,angle=0]{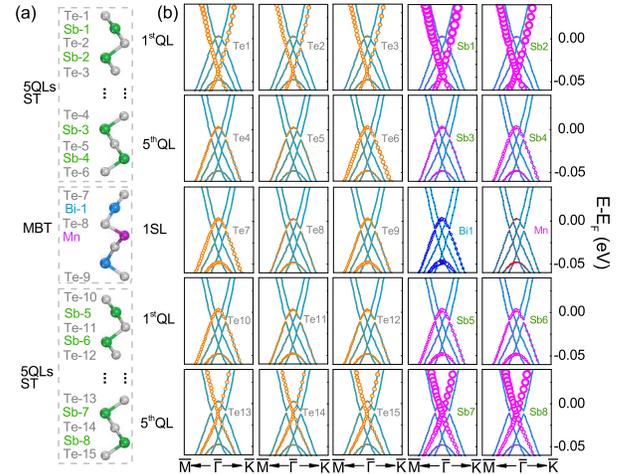}
  \caption{(a) The crystal structure and label numbers of some important atoms in the ST/MBT/ST heterostructure. (b) The characters band near the Fermi level for ST/MBT/ST heterostructure by projecting Kohn-Sham states to the local orbitals of a single atom. The contribution of each atom is charactered by the size of circles in different colors.} \label{Fig4}
\end{figure}

Why is the Dirac point formed in the ST/MBT/ST heterostructure? Based above discussion, the surface carrier mediated FM can penetrate into 2-3 QLs of the first and second ST TIs adjent to MBT. But the distance between top layer of the first ST (5QLs) and bottom layer of the second ST(5QLs) is larger than penetration depths of the bottom layer of the first ST and top layer of the second layer. Hence, long-range FM coupling between top layer of the first ST and bottom layer of the second ST can not be realized. Based on above analysis, the ST/MBT/ST system should be still a TI one. Thus, the Dirac point is roboust for the ST/MBT/ST heterostructure. We also provide the detailed atomic contribution to the Dirac point of the ST/MBT/ST system. As shown in the Figs.~\ref{Fig4}, the bands around the Dirac point are mainly contributed by the atoms from the bottom QL of the second ST, which is in agreement with above analysis based on FM coupling of the ST/MBT/ST system. In addition, we also calculated the Z2 value of the ST/MBT/ST heterostructure and obtained Z2=1. It confirms that the ST/MBT/ST is a topological nontrivial system.

\begin{figure}
  \includegraphics[width=8cm,angle=0]{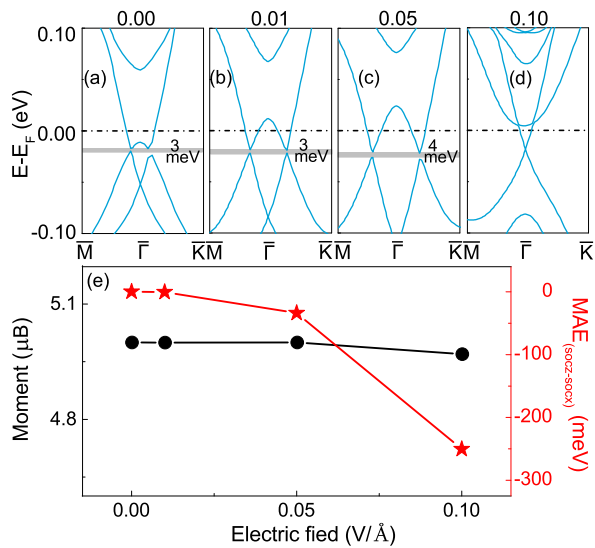}
  \caption{(a)-(d)Band structures, (e) total magnetic moments and magnetocrystalline anisotropy energy of MBT/ST heterostructure under zero electric field and electric field (Direction of electric-field strength E is from Sb$_2$Te$_3$ to MnBi$_2$Te$_4$, which is same to that of the built-in electric field of the system due to MnBi$_2$Te$_4$ with intrinsic n-type defects and Sb$_2$Te$_3$ with intrinsic p-type defects).} \label{Fig5}
\end{figure}

In experiments, the fabricated MnBi$_2$Te$_4$ has n-type defects and Sb$_2$Te$_3$ has p-type defects, which might induce a built-in electric field in MnBi$_2$Te$_4$/Sb$_2$Te$_3$ heterostructure. We have performed some calculations to estimate the effect of built-in electric field on the band structure and magnetic property of MnBi$_2$Te$_4$/Sb$_2$Te$_3$ heterostructure. As shown in the Figs.~\ref{Fig5}(a)-(d), in comparsion with the band structure without built-in field, the band gap changes quite slightly when a small electric field is applied, but it will lead to gap closed when electric field reaches to 0.1 V/\AA. From the Figs.~\ref{Fig5}(e), we can obtain that electric field almost has not effect on magnetic moment of the MBT/ST system, but it will obviously make the out-of-plane magnetic structure more stable than that of in-plane magnetic structure of MBT/ST system. Hence, low concentration of intrinsic defects in Sb$_2$Te$_3$ and MnBi$_2$Te$_4$ will have neglected influence on the band gaps, but may enhance magnetocrystalline anisotropy energy of MBT/ST heterostructures.

\begin{figure}
  \includegraphics[width=8cm,angle=0]{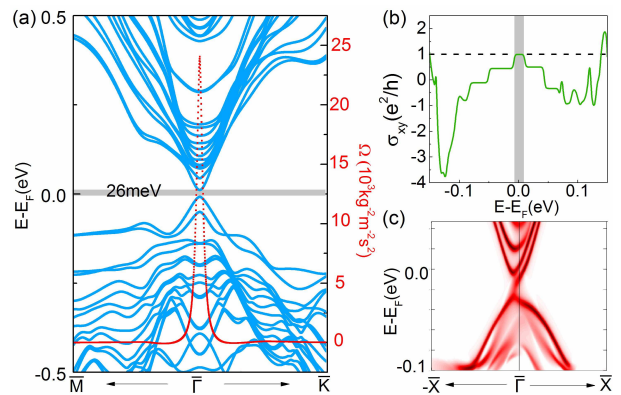}
  \caption{Band structures along high-symmetry lines (a) and near the $\Gamma$ point (b) of MBT/ST/MBT heterostructure. The corresponding Berry curvatures are indicated in the (a). (c) Hall conductivity as a function of energy. (d) Band structures of semi-infinite systems including the chiral edge state.
  } \label{Fig6}
\end{figure}

We now verify that the opened band gap in MBT/ST/MBT system hosts a topologically nontrivial insulator, i.e., QAHE, by calculating the Chern number, which can be obtained by integrating the Berry curvature of the occupied valence bands~\cite{BerryCurvature1,BerryCurvature2}.
Figure~\ref{Fig6}(a) displays the Berry curvature distribution along the high symmetry lines, where a large negative peak appears near the $\Gamma$ point and zero elsewhere. As a consequence, the total integration or the Hall conductance with the Fermi level lying inside the band gap must be nonzero. The nonzero Berry curvature means that the electrons flow in the curl field and contribute to the quantized Hall conductivity. Figure~\ref{Fig6}(b) displays the Hall conductance $\sigma_{xy}$ as a function of the energy. One can clearly see that $\sigma_{xy}=+e^{2}/h$ in the energy interval near Fermi level. Hence, the nontrivial band gap for MBT/ST/MBT is 26 meV, which is beneficial for realizing high-temperature QAHE. The chiral edge states are also provided by using the surface Green function method as implemented in the WannierTools~\cite{WannierTools}. For MBT/ST/MBT system, one edge state [see Fig.~\ref{Fig6}(c)] emerges inside the band gap, connecting the valence and conduction bands, and corresponding to the Chern number of $\mathcal{C}=1$. How about the QAHE observation temperature in MBT/ST/MBT system? Based on the previously obtained Curie temperature of $T_{\rm C}=42 K$ and the gap size of 26 meV, one can conclude that the QAHE observation temperature can reach up to 42 K.

\textit{Summary---.}
In conclusion, our findings report that a natural magnetic van der Waals heterostructures composed of MnBi$_2$Te$_4$ and Sb$_2$Te$_3$ can be constructed to realize high-temperature QAHE. We provide proof-of-principle numerical demonstration that: (1) an interlayer ferromagnetic phase transition can occur at $T_{\rm C}$=42K when a 5QL Sb$_2$Te$_3$ is inserted into two SL MnBi$_2$Te$_4$ with interlayer antiferromagnetic coupling; (2) Band structures and topological property calculations verify that MnBi$_2$Te$_4$/Sb$_2$Te$_3$/MnBi$_2$Te$_4$ heterostructure system is an intrinsic QAHE with a surface gap ~26 meV and the corresponding Chern number is $\mathcal{C}=1$. In addition, considering the charge compensation fact due to intrinsic $n$-type defects in MnBi$_2$Te$_4$ and $p$-type defects in Sb$_2$Te$_3$ in experiments, MnBi$_2$Te$_4$/Sb$_2$Te$_3$/MnBi$_2$Te$_4$ system should be an very appropriate candidate for exploring high-temperature QAHE.

\begin{acknowledgments}
This work was financially supported by the National Key R\&D Program (2017YFB0405703), NNSFC (11974098, 61434002 and 11474265), Natural Science Foundation of Hebei Province (A2019205037), Science Foundation of Hebei Normal University (2019B08), Anhui Initiative in Quantum Information Technologies. We are grateful to AMHPC and Supercomputing Center of USTC for providing the high-performance computing resources.
\end{acknowledgments}

\end{document}